\documentclass[aps,prl,showpacs,twocolumn,nofootinbib,10pt,showkeys]{revtex4-1}
\usepackage{epsfig,epsf,graphics}
\usepackage{epstopdf}
\usepackage{color}
\usepackage{amsfonts,amssymb,stmaryrd,latexsym,amsmath}
\usepackage{hhline}

\begin{document}
\title{Consequences of Heavy Quark Symmetries for Hadronic Molecules}
\author{Feng-Kun Guo}
\email{fkguo@hiskp.uni-bonn.de}
\affiliation{Helmholtz-Institut f\"ur Strahlen- und
             Kernphysik and Bethe Center for Theoretical Physics, \\
             Universit\"at Bonn,  D-53115 Bonn, Germany}
\author{Carlos Hidalgo-Duque}
\email{carloshd@ific.uv.es}
\affiliation{Instituto de F\'isica Corpuscular (IFIC),
             Centro Mixto CSIC-Universidad de Valencia,
             Institutos de Investigaci\'on de Paterna,
             Aptd. 22085, E-46071 Valencia, Spain}
\author{Juan Nieves}
\email{jmnieves@ific.uv.es}
\affiliation{Instituto de F\'isica Corpuscular (IFIC),
             Centro Mixto CSIC-Universidad de Valencia,
             Institutos de Investigaci\'on de Paterna,
             Aptd. 22085, E-46071 Valencia, Spain}
\author{Manuel Pav\'on Valderrama}
\email{pavonvalderrama@ipno.in2p3.fr}
\affiliation{Institut de Physique Nucl\'eaire,
             Universit\'e Paris-Sud, IN2P3/CNRS,
             F-91406 Orsay Cedex, France}

\begin{abstract}
Among the newly observed structures in the heavy quarkonium mass region,
some have been proposed to be hadronic molecules.
We investigate the consequences of heavy quark flavor symmetry
on these heavy meson hadronic molecules.
The symmetry allows us to predict new hadronic molecules
on one hand, and test the hadronic molecular assumption of
the observed structures on the other hand.
We explore the consequences of the flavor symmetry assuming
the $X(3872)$ and $Z_b(10610)$ as a isoscalar $D\bar D^*$
and isovector $B\bar B^*$ hadronic molecule, respectively.
A series of hadronic molecules composed of heavy mesons are predicted.
In particular, there is an isoscalar $1^{++}$ $B\bar B^*$ bound state
with a mass about $10580$~MeV which may be searched for in the
$\Upsilon(1S,2S)\pi^+\pi^-\pi^0$ mass distribution; the isovector
charmonium partners of the $Z_b(10610)$ and the $Z_b(10650)$ are also
predicted, which probably corresponds to the very recently
observed $Z_c(3900)$ and $Z_c(4025)$ resonances
by the BESIII Collaboration.
\end{abstract}


\maketitle

\vspace{1cm}

Due to ongoing experimental efforts, a series of new heavy quarkonium states,
called $XYZ$ states, have been observed in the last decade.
Many of them are expected to be of exotic nature,
for a comprehensive review we refer to Ref.~\cite{Brambilla:2010cs}.
Due to the proximity of the masses to certain hadronic thresholds,
some of the $XYZ$ states have been proposed
to be hadronic molecules, i.e. states that are generated by
the interaction between two or more hadrons (they
are bound states if they are below the threshold
and in the first Riemann sheet, or virtual states
and resonances if they are in the second
Riemann sheet of the scattering amplitude).
For instance, the famous $X(3872)$ discovered by
the Belle Collaboration~\cite{Choi:2003ue} and confirmed by many
other experiments was proposed to be a $D\bar D^*$ (the charge conjugated
particles are implicitly included here and in the following) bound state with an
extremely small binding energy~\cite{Tornqvist:2004qy} or a virtual
state~\cite{Hanhart:2007yq}; the isovector $Z_b^{\pm,0}(10610)$ and
$Z_b^{\pm}(10650)$ resonances reported by the Belle
Collaboration~\cite{Belle:2011aa,Adachi:2012im} have been considered
to be $B\bar B^*$ and $B^*\bar B^*$ hadronic molecules, respectively~\cite{Bondar:2011ev};
the $Y(4660)$ observed in the $\psi'\pi\pi$ mass distribution by the Belle
Collaboration~\cite{Wang:2007ea} and confirmed by BaBar~\cite{Lees:2012pv}
is possibly a $\psi' f_0(980)$ bound state~\cite{Guo:2008zg};
the $X(4260)$ state observed by the BaBar
Collaboration~\cite{Aubert:2005rm} has been suggested to be a $D\bar D_1$
molecule~\cite{Ding:2008gr,Li:2013bca,Wang:2013cya}.
Other models also exist for these states.
Thus, in order to understand these states and furthermore their binding
mechanisms, it is important to find out methods which can distinguish the
different scenarios. Decay patterns are often used for this purpose, here we
will pursue a different approach using heavy flavor symmetry.

Without developing complicated dynamical models, symmetries are often helpful in
describing certain aspects of various physical systems. For a system involving
a heavy quark whose mass $m_Q$ is much larger than $\Lambda_\text{QCD}$, flavor
and spin symmetries arise by sending $m_Q$ to infinity (for reviews of heavy
quark symmetries, see Refs.~\cite{Neubert:1993mb,manoharbook}). Due to spin
symmetry, both heavy mesons and heavy quarkonia form spin multiplets, e.g. the
$\{D,D^*\}$ and $\{\eta_c,J/\psi\}$. The masses are degenerate in the heavy
quark limit, and their interactions with other hadrons are the same at leading
order (LO).
Heavy quark spin symmetry was already widely used in predicting new hadronic
molecules~\cite{Guo:2009id,Bondar:2011ev,Voloshin:2011qa,Mehen:2011yh,Nieves:2011zz,Valderrama:2012jv,Nieves:2012tt,HidalgoDuque:2012pq}.
In this work, in addition to spin symmetry, we will argue that heavy quark
flavor symmetry is also very useful in the context of hadronic molecules, and
may be used to verify hadronic molecular hypothesis and predict new hadronic
molecules. As examples, assuming the $X(3872)$ and $Z_b(10610)$ to
be the $D\bar D^*$ and $B\bar B^*$ molecules, we can predict a
series of new hadronic molecules, including the $Z_c(3900)$ reported
very recently by the BESIII Collaboration~\cite{BESIII:2013},
later also by Belle~\cite{Liu:2013xoa}, and the new $Z_c(4025)$
observed by BESIII~\cite{Ablikim:2013emm}.

Let us consider the interaction between two heavy hadrons
forming a bound state.
As far as the hadrons are not too tightly bound, they will not probe
the specific details of the interaction binding them at short distances.
Moreover, each of the constituent heavy hadrons will be unable
to {\it see} the internal structure of the other heavy hadron.
Therefore, we can exploit this separation of scales to formulate
an effective field theory (EFT) description of hadronic molecules.
Within EFT we express physical quantities as power series in terms
of the ratio $Q / M$, where $Q$ stands for the momenta of the
mesons within a molecule or the pion mass and $M$ is the QCD
hadronic mass scale (of the order of the $\rho$ mass or the
center of mass momentum necessary for a heavy hadron
to excite another).
The contribution of physics at the hard scale $M$ is safely encoded
in the counterterms of the EFT at low energies~\cite{Ecker:1988te, Epelbaum:2001fm}.

The situation is analogous to that of the EFT formulation of
the nucleon-nucleon interaction~\cite{Epelbaum:2008ga},
which we use as a template for the EFT of heavy hadronic molecules.
Yet heavy hadrons entail interesting simplifications over nucleons.
On the one hand heavy quark symmetry heavily constrains
the low-energy interactions among heavy
hadrons~\cite{AlFiky:2005jd,Mehen:2011yh}.
On the other hand pion exchanges are in general 
perturbative~\cite{Fleming:2007rp,Valderrama:2012jv},
in contrast to nuclear physics where they are not~\cite{Birse:2005um},
and produce small effects.
The only exception is the isoscalar bottom sector where the pions 
might be nonperturbative due to the large masses of bottom 
mesons~\cite{Valderrama:2012jv,Mehen:2011yh}. Yet, the inclusion of one pion 
exchange in this sector only introduces minor modifications of the predictions, 
c.f. the discussion of the numerical results later on~\footnote{Because the isospin 
factor in the isovector case is only 1/3 of that in the isoscalar case, the 
pions are perturbative again in the isovector bottom sector.}.
As a consequence, at ${\rm LO}$ the EFT description only involves
energy-independent contact range
interactions~\cite{Valderrama:2012jv,Nieves:2012tt}.

We are mainly interested in two manifestations of heavy quark symmetry:
heavy quark spin symmetry (HQSS) and heavy flavour symmetry (HFS).
Their role can be easily illustrated in the heavy meson-antimeson system
with a series of examples.
We begin with HQSS as applied to the $Z_b$ and $Z_b'$, where we assume
that they are $1^{+-}$ $B\bar{B}^*$ and $B^*\bar{B}^*$ isovector
molecules, respectively.
HQSS implies that the ${\rm LO}$ non-relativistic isovector
heavy meson-antimeson potential is identical
in both cases~\cite{Bondar:2011ev,Voloshin:2011qa,Mehen:2011yh}
\begin{eqnarray}
V^{\rm LO}_{B\bar{B}^*}(1^{+-}) = V^{\rm LO}_{B^*\bar{B}^*}(1^{+-}) \, ,
\end{eqnarray}
where we have indicated the particle channel in the subscript.
This explains why the energy shift of the $Z_b$/$Z_b'$ states relative
to the $B\bar{B}^*$/$B^*\bar{B}^*$ thresholds is almost the same.
For a further example we can consider the $X(3872)$
--- the $X_c$ from now on --- as a $1^{++}$ $D\bar{D}^*$ molecule.
HQSS then predicts that the potential in the $X_c$ channel is the identical
to that of the $2^{++}$ $D^*\bar{D}^*$ channel~\cite{Valderrama:2012jv,Nieves:2012tt}:
\begin{eqnarray}
V^{\rm LO}_{D\bar{D}^*}(1^{++}) = V^{\rm LO}_{D^*\bar{D}^*}(2^{++}) \, ,
\end{eqnarray}
meaning that we can expect the existence of a $2^{++}$
HQSS partner of the $X(3872)$.
Explicit calculations indicate that its mass should be in the vicinity
of $4012\,{\rm MeV}$~\cite{Nieves:2012tt}.
Following the previous naming pattern, we will call this state the $X_c'$.

As can be appreciated the exciting feature about heavy meson molecules is
their high degree of symmetry.
This is even more evident when we consider HFS.
According to HFS the interactions involving heavy mesons do not depend
on the heavy quark flavour.
This means that the heavy meson-antimeson potential is not able
to distinguish the $D$/$D^*$ mesons from the $B$/$B^*$ ones.
If we apply this idea to the $X_c$, we find
\begin{eqnarray}
V^{\rm LO}_{D\bar{D}^*}(1^{++}) = V^{\rm LO}_{B\bar{B}^*}(1^{++}) \, ,
\end{eqnarray}
and the same is true for the potentials
in the $X_c'$, $Z_b$ and $Z_b'$ channels.
The consequence of HFS is that heavy meson molecules
can appear in flavour multiplets.
A resonance in the charm sector
might have a counterpart in the bottom sector and vice versa.
However, there is a catch.
The formation of bound states does not only depend on the strength of
the potential, but also on the reduced mass of the two-body system.
A higher reduced mass translates into a stronger binding.
If the $X_c$ binds, it is more than likely that the $X_b$ --- the bottom
counterpart of the $X_c$ --- binds too.
Searching for such a state may even be regarded as a test of the hadronic
molecular hypothesis of the $X(3872)$.
On the contrary, the shallow nature of the $Z_b$ and $Z_b'$ indicates
that their charm counterparts are probably unbound.
Yet the $Z_c$ and $Z_c'$ might survive as virtual states or resonances.
As we will see, this is indeed the case.

Now we compute the the expected location of the HQSS and HFS partners
of the $X_c$, $Z_b$ and $Z_b'$.
For that, we notice that at ${\rm LO}$ the EFT potential is simply
a contact-range interaction of the type
\begin{eqnarray}
\langle \vec{p}\, | V_{X}^{\rm LO} | \vec{p}\,' \, \rangle &=& C_{0X} \, , \\
\langle \vec{p}\, | V_{Z}^{\rm LO} | \vec{p}\,' \, \rangle &=& C_{1Z} \, ,
\end{eqnarray}
where the subscripts indicate the isospin and whether we are considering
an $X$- or $Z$-like channel (see Table \ref{tab:HHstates}).
For finding bound state solutions we iterate this potential
in the Lippmann-Schwinger equation (LSE), where the details
can be consulted in Ref.~\cite{Nieves:2012tt}.
At this point we find it worth commenting that the contact-range potential
is singular and requires a regularization and renormalization
procedure. 
We employ a standard gaussian regulator with a cut-off
$\Lambda = 0.5-1\,{\rm GeV}$, where we have chosen
the cut-off window according to the following
principles: $\Lambda$ must be bigger than the wave number
of the states, but at the same time  must be small enough to
preserve heavy quark symmetry and prevent that the theory might
become sensitive to the specific details of short distance dynamics.
The dependence of results on the cut-off, when it varies within this window,
provides an estimate of the expected size of subleading corrections.
For a more complete discussion on the choice of the cut-off in
nucleon-nucleon systems, see for instance Ref.~\cite{Epelbaum:2009sd}.

\begin{table}[tb]
\caption{\label{tab:HHstates} Various combinations having
   the same contact term as the $X(3872)$ (left) and $Z_b(10610)$ (right). Here
   $P$ and $P^*$ represent $D,\bar B$ and $D^*,\bar B^*$, respectively.}
\begin{ruledtabular}
\begin{tabular}{l c | l c}
       $I(J^{PC})$ & $C_{0X}$ & $I(J^{PC})$ & $C_{1Z}$ \\
      \hline
      $0(1^{++})$ & $\frac1{\sqrt{2}}\left(P\bar P^*-P^*\bar P\right)$ &
      $1(1^{+-})$ & $\frac1{\sqrt{2}}\left(P\bar P^*+P^*\bar P\right)$ \\
      $0(2^{++})$ & $P^*\bar P^*$ & $1(1^{+-})$ & $P^*\bar P^*$ \\
      $0(2^{+})$ & $D^*B^*$ & $1(1^{+})$ & $D^*B^*$  \\
   \end{tabular}
\end{ruledtabular}
\end{table}

\begin{table*}[tb]
\caption{\label{tab:predictions} Heavy meson--heavy meson combinations
  having the same contact term as the $X(3872)$ and $Z_b(10610)$, and
  the predictions of the pole positions, which are understood to
  correspond to bound states except if we write ``V'' in parenthesis
  for denoting a virtual state. When we increase the strength of the
  potential to account for the various uncertainties, in one case
  (marked with $\dagger$ in the table) the virtual pole
  evolves into a bound state. The masses are given in units of
  MeV.}
\begin{ruledtabular}
\begin{tabular}{l c c c c c c}
       $V_C$ & $I(J^{PC})$ & States & Thresholds & Masses ($\Lambda=0.5$ GeV) &
       Masses ($\Lambda=1$ GeV) & Measurements
       \\\hline
       $C_{0X}$ & $0(1^{++})$ & $\frac1{\sqrt{2}}(D\bar D^*-D^*\bar D)$ &
       3875.87 & 3871.68 (input) &  3871.68 (input) &
       $3871.68\pm0.17$~\cite{PDG}  \\
                       & $0(2^{++})$ & $D^*\bar D^*$ &
       4017.3  & $4012^{+4}_{-5}$  &  $4012^{+5}_{-12}$ & ?\\
       & $0(1^{++})$ & $\frac1{\sqrt{2}}(B\bar B^*-B^*\bar B)$ &
       10604.4 & $10580^{+9}_{-8}$ &  $10539^{+25}_{-27}$ & ?\\
                       & $0(2^{++})$ & $B^*\bar B^*$ &
       10650.2 & $10626^{+8}_{-9}$ & $10584^{+25}_{-27}$ & ?\\
                       & $0(2^{+})$ & $D^*B^*$ &
       7333.7 & $7322^{+6}_{-7}$ & $7308^{+16}_{-20}$ & ?\\ \hline
       $C_{0Z}$ & $1(1^{+-})$ & $\frac1{\sqrt{2}}(B\bar B^*+B^*\bar B)$ &
        10604.4 & $10602.4 \pm 2.0$ (input) & $10602.4 \pm 2.0$ (input) &
       $10607.2\pm2.0$~\cite{Belle:2011aa} \\
       & & & & & & $10597\pm9$~\cite{Adachi:2012cx}\\
                       & $1(1^{+-})$ & $B^*\bar B^*$ &
  10650.2 & $10648.1 \pm 2.1 $ &  $10648.1 ^{+2.1}_{-2.5}$ & $10652.2\pm1.5$~\cite{Belle:2011aa} \\
       & & & & & & $10649\pm12$~\cite{Adachi:2012cx}\\
                       & $1(1^{+-})$ & $\frac1{\sqrt{2}}(D\bar D^*+D^*\bar D)$ &
       3875.87 & $3871^{+4}_{-12}$ (V) &  $3837_{-35}^{+17}$ (V) & $3899.0 \pm 3.6 \pm 4.9$~\cite{BESIII:2013} \\
       & & & & & & $3894.5 \pm 6.6 \pm 4.5$~\cite{Liu:2013xoa}\\
                       & $1(1^{+-})$ & $D^*\bar D^*$ &
       4017.3 & $4013^{+4}_{-11}$ (V) &  $3983_{-32}^{+17}$ (V) & $4026.3 \pm 2.6$~\cite{Ablikim:2013emm}. \\
                       & $1(1^{+})$ & $D^*B^*$ &
       7333.7 & $7333.6^{\dagger}_{-4.2}$ (V) & $7328^{+5}_{-14}$ (V) & ?\\
   \end{tabular}
\end{ruledtabular}
\end{table*}

For the numerical calculations, we work in the isospin symmetric limit
and use the averaged masses of the heavy mesons, which are
$M_D = 1867.24$~MeV, $M_{D^*} = 2008.63$~MeV, $M_B =
5279.34$~MeV and $M_{B^*} = 5325.1$~MeV.
The value of $C_{0X}$ is determined from reproducing the central value of
the Particle Data Group averaged mass of the $X_c(3872)$,
$3871.68\pm0.17$~GeV~\cite{PDG}.
The resulting value is $C_{0X} = -1.94~\text{fm}^2$ for $\Lambda=0.5$~GeV
and $-0.79~\text{fm}^2$ for $\Lambda=1$~GeV~\cite{Nieves:2012tt},
where the uncertainties coming from the error in the mass of
the $X_c$ are negligible.
At this point one may argue that isospin breaking is important for the
$X_c$, owing to its closeness to the $D^0 \bar{D}^{0*}$ threshold,
but concrete calculations indicate that the effect
is tiny for spectroscopy~\cite{HidalgoDuque:2012pq}.
In turn, the value of $C_{1Z}$ may be fixed using the $Z_b(10610)$ mass.
The mass of the $Z_b(10610)$ measured in the $\Upsilon(nS)\pi,h_b(nP)\pi$
distribution $10607.2\pm2.0$~MeV~\cite{Belle:2011aa} is $1.3\,\sigma$
above the $B \bar B^*$ threshold,
while the value measured in the $\Upsilon(5S)\to B\bar B^*\pi$ decay
$10597\pm9$~MeV~\cite{Adachi:2012cx} overlaps with the $B\bar B^*$ threshold.
However, these estimations are based on parametrizing the $Z_b$ and $Z_b'$
poles as Breit-Wigner. The analysis of Ref.~\cite{Cleven:2011gp}, which
overcomes this limitation, suggests that the $Z_b$ and $Z_b'$ are
slightly below threshold and have a binding energy of
$\sim 4.7\,{\rm MeV}$ and $\sim 0.1\,{\rm MeV}$
respectively.
In line with the estimates of Ref.~\cite{Cleven:2011gp},
we assume the $Z_b$ binding energy to be $2.0 \pm 2.0\,{\rm MeV}$,
yielding $C_{1Z} = -0.75^{+15}_{-28}~\text{fm}^2$  for $\Lambda=0.5$~GeV and
$-0.30^{+3}_{-7}~\text{fm}^2$ for $\Lambda=1$~GeV.

With these values we can make predictions by solving the LSE, as previously
commented.
We summarize our results in Table~\ref{tab:predictions}.
The uncertainties that are listed correspond to taking into account that
HQSS and HFS are not exact, but approximate.
We expect a ${\Lambda_{\rm QCD}}/{m_Q}$ deviation of the $C_{0X}$ and
$C_{1Z}$ value from the heavy quark limit. Taking 300~MeV for
$\Lambda_\text{QCD}$~\cite{PDG},  and 1.5~GeV and 4.5 GeV for $m_c$ and $m_b$, 
respectively, this translates into a relative $20\%$ error in the charm sector 
and $7\%$ in the bottom one.
Actually, the errors are dominated by the uncertainty in the charm sector.
When we compute the $X_b$ and $X_b'$, the relative error of $C_{0X}$ is
rather $20\%$ than $7\%$ as its value has been determined
from the $X_c(3872)$.
We remind that the uncertainties coming from the errors in the mass
of the $X_c(3872)$ are negligible in comparison.
For the states derived from the $Z_b$'s we sum the ${\Lambda_{\rm QCD}}/{m_Q}$
and the binding energy errors in quadrature,
where the binding error dominates. 
Some of the states --- the partners of the $Z_b$/$Z_b'$ --- are not bound,
but virtual. We indicate this with a ``V''.

Among the predicted states, the $2^{++}$ ones can decay into two heavy
pseudoscalar mesons in a $D$ wave, which would introduce a width of order
$\mathcal{O}(10~\text{MeV})$.
We refer the refined results taking into account
the coupled channels to a forthcoming work.
The predicted mass of the $D^*\bar D^*$ bound state is higher than
the $\chi_{c2}(2P)$ with a mass of $3927.2\pm2.6$~MeV~\cite{PDG},
and might be searched for in the same process as
the $\chi_{c2}(2P)$, i.e. $\gamma\gamma\to D\bar D$.
The data collected at both Belle and BaBar~\cite{PRL96.082003,PRD81.092003}
in that range do not have enough statistics for concluding
the existence of such a state.

The most robust prediction would be the $B\bar B^*$ bound state
with $I(J^{PC})=0(1^{++})$, to be called $X_b(10580)$,
the analogue of the $X_c(3872)$ in the bottom sector.
As mentioned earlier, the pions in  this sector might be
nonperturbative. One may worry about the stability of the results in this 
sector against including the pions. 
However, we have checked that one pion exchange only slightly changes the 
central value of the $X_b$ mass to 10584~MeV and 10567~MeV for $\Lambda=0.5$~GeV 
and 1~GeV, respectively (notice that the cut-off
dependence decreases). 
This state should be narrow since the decay into the $B\bar B$ is forbidden.
It would decay dominantly into a bottomonium and light mesons.
Moreover, the difference between the charged and neutral $B\bar B^*$ threshold
is tiny, and completely negligible when compared with the binding energy.
Therefore, unlike the $X_c(3872)$,
whose decays exhibit a large isospin breaking,
the $X_b(10580)$ would decay into $\Upsilon(nS)\pi\pi\pi$ ($n=1,2$) rather
than $\Upsilon(nS)\pi\pi$. 
It can also decay into $\chi_{bJ}(nP)$ and pions.
It is worth emphasizing that the existence of such a state is a 
consequence of HFS and the assumption of the $X_c(3872)$ being a $ D\bar D^*$ 
bound state. Searching for it would shed light on the nature of the 
$X_c(3872)$.

The $Z_c$ and $Z_c'$ appear as virtual states, not very far away
from their respective thresholds.
However, the uncertainties of the ${\rm LO}$ calculation are large,
of the order of tens of MeV,  as indicated by the difference 
between the results with different cut-off values. 
From this point of view, 
the new charged structure $Z_c(3900)$ observed by the BESIII 
Collaboration~\cite{BESIII:2013}, and confirmed by the Belle 
Collaboration~\cite{Liu:2013xoa} and an analysis using the CLEO 
data~\cite{Xiao:2013iha}, is a natural candidate for the partner
in the charm  sector of the $Z_b(10610)$. 
Analogously, we expect the recent $Z_c(4025)$~\cite{Ablikim:2013emm}
to be the partner of the $Z_b(10650)$.
Therefore, we are tempted to identify the $Z_c(3870)$ and 
$Z_c(4010)$ states reported in Table~\ref{tab:predictions}
with the observed $Z_c(3900)$ and $Z_c(4025)$.
We observe that the $Z_c$ and $Z_c'$ are not necessarily virtual:
there are subleading order dynamics that can easily
move the states above threshold.
Most notably at next-to-leading order the EFT potential can develop
a short range repulsive barrier.
Thus the LO uncertainty also encompasses the possibility
that the states might be resonant.
There are also corrections coming from coupled channel dynamics,
but in general they are at least next-to-next-to-leading order
and hence their impact is modest at best.
For instance, the $Z_c$ and $Z_c'$ channels couple with each other
and with the nearby $h_c(2P)\pi$ and $\psi(2S)\pi$ channels,
though in the latter case we do not know the location of
these charmonia.
Their impact could be enhanced if they are close enough to the $Z_c$/$Z_c'$
poles (yet they will continue to be subleading).
All this indicates that the $Z_c$ and $Z_c'$ are promising candidates
to explain the recently observed $Z_c(3900)$ and $Z_c(4025)$ resonances,
though further theoretical effort is still required.

To summarize, in this work we have argued that in addition to HQSS, 
HFS can be used to predict new heavy meson molecules.
We have also considered the uncertainties due to the finite mass of
the heavy quarks.
The predictions are important in understanding the newly observed hadrons
in the heavy quarkonium mass region in the sense that, 
if the $XYZ$ states are hadronic molecules, they will probably
have heavy flavour partners that should be searched for.
Note that HFS is a symmetry among the coefficients in the interaction of
the Lagrangians (or, equivalently, the heavy meson potentials),
and not a symmetry in the binding energies (as the kinetic
term of the Lagrangian breaks the symmetry):
what matters for binding is the potential times the reduced mass
of the heavy hadrons.
Particularly, we studied in detail the new states that can be derived from
the hypothesis that the $X(3872)$ and $Z_b(10610)$ are
$D\bar D^*$ and $B\bar B^*$ hadronic molecules,
respectively~\footnote{We notice that approaches involving
phenomenological (i.e. model-dependent) ingredients -- but
usually incorporating heavy quark symmetry --
can lead to other conclusions:
while the $X_b$ is usually predicted,
the $Z_c$ is not~\cite{Tornqvist:1993ng,Sun:2011uh,Nieves:2011zz}.}.
Searching for the isoscalar $1^{++}$ $B\bar B^*$ bound state in the
$\Upsilon(1S,2S)\pi^+\pi^-\pi^0$ channel at hadron colliders or photon-photon
collisions would provide valuable information on the structure of the $X(3872)$.
In addition, we find promising isovector $1^{+-}$ $D\bar D^*$
and $D^*\bar D^*$ virtual states near threshold
that could very well be identified with the newly discovered
$Z_c(3900)$~\cite{BESIII:2013} and $Z_c(4025)$~\cite{Ablikim:2013emm}.

\medskip

We would like to thank Qiang Zhao for useful discussions. This work
was initialized during the Bethe Forum on Exotic
Hadrons. F.-K.G. acknowledges the Theory Division of IHEP in Beijing,
where part of the work was done, for the hospitality. C. H.-D. thanks
the support of the JAE-CSIC Program. This work is supported in part by
the DFG and the NSFC through funds provided to the Sino-German CRC 110
``Symmetries and the Emergence of Structure in QCD'', by the NSFC
(Grant No. 11165005), by the Spanish Ministerio de Econom\'\i a y
Competitividad and European FEDER funds under the contract
FIS2011-28853-C02-02 and the Spanish Consolider-Ingenio 2010 Programme
CPAN (CSD2007-00042), by Generalitat Valenciana under contract
PROMETEO/2009/0090 and by the EU HadronPhysics2 project, grant
agreement no. 227431.


\end{document}